\documentclass[aip,rsi,amsmath,amssymb,reprint]{revtex4-1}

\usepackage{graphicx}
\usepackage{dcolumn}
\usepackage{bm}
\usepackage{physics}
\usepackage{siunitx}
\usepackage{epstopdf}
\usepackage{gensymb}
\usepackage{hyperref}
\usepackage{color}

\usepackage{natbib}
\begin{document}

\preprint{AIP/123-QED}

\title[Sample title]{A low-steering piezo-driven mirror}

\author{E. Magnan}
\altaffiliation[Also at]{ Laboratoire Charles Fabry, Institut d'Optique Graduate School,
CNRS, Universit\'e Paris-Saclay, 91127 Palaiseau cedex, France.}

\author{J. Maslek}
\affiliation{Joint Quantum Institute, National Institute of Standards and Technology
and the University of Maryland, College Park, Maryland 20742 USA}

\author{C. Bracamontes}
\affiliation{Joint Quantum Institute, National Institute of Standards and Technology
and the University of Maryland, College Park, Maryland 20742 USA}

\author{A. Restelli}
\affiliation{Joint Quantum Institute, National Institute of Standards and Technology
and the University of Maryland, College Park, Maryland 20742 USA}

\author{T. Boulier}
\altaffiliation[Also at]{ Laboratoire Charles Fabry, Institut d'Optique Graduate School,
CNRS, Universit\'e Paris-Saclay, 91127 Palaiseau cedex, France.}

\author{J.V. Porto}
\affiliation{Joint Quantum Institute, National Institute of Standards and Technology
and the University of Maryland, College Park, Maryland 20742 USA}

\date{\today}


\begin{abstract}
We present a piezo-driven translatable mirror with excellent pointing stability, capable of driving at frequencies up to tens of kilohertz. Our system uses a tripod of piezo actuators with independently controllable drive voltages, where the ratios of the individual drive voltages are tuned to minimize residual tilting. Attached to a standard $\varnothing=\SI{12.7}{\milli\meter}$ mirror, the system has a resonance-free mechanical bandwidth up to $\SI{51}{\kilo\hertz}$, with displacements up to $\SI{2}{\micro\meter}$ at $\SI{8}{\kilo\hertz}$. The maximum static steering error is $\SI{5.5}{\micro\radian}$ per micrometer displaced and the dynamic steering error is lower than $\SI{0.6}{\micro\radian\micro\meter}^{-1}$. This simple design should be useful for a large set of optical applications where tilt-free displacements are required, and we demonstrate its application in an ensemble of cold atoms trapped in periodically driven optical lattices.
\end{abstract}

\pacs{Valid PACS appear here}
\maketitle


\section{Introduction}
Piezo-based opto-mechanical devices, such as tip-tilt stages, are routinely used for applications as diverse as image stabilization\cite{Eromaki2007,1990spie}, adaptive optics\cite{Roddier1999,Wang2017,Davies2012}, microscopy\cite{Wolleschensky2004,Salapaka2002}, optical communication systems\cite{Gorman2003} and laser stabilization\cite{Hall2010}. The typical figures of merit for such devices are the modulation bandwidth of the mechanical displacement, and the maximum possible displacement. For some applications, minimal angular rotation of the optical element while it is translated is essential, but typical designs\cite{Chadi2013} fail to compensate for undesired tilts when driving the piezo. This is particularly true if the optical path lengths in the system are long, such that angular rotation leads to large beam displacements. One approach to minimize beam steering is to stabilize rotations using kinematic restrictions that allow for translation but not rotation. The kinematic approach has the drawback that it can be complicated to implement, and the increased mass and friction of such a design can limit the bandwidth of the modulation response. We present here a mechanically simple design that provides low rotation during  translation, while maintaining large mechanical bandwidths.\\
\\
\indent Ultra-cold neutral atoms can be optically trapped in the interference patterns of light to form so called ``optical lattices''\cite{Grimm2000,Bloch2005a}, which allows for cold atom simulation of crystalline many-body physics. The most common way to create such an optical lattice is to retro-reflect a laser beam from a mirror, where the position of the optical lattice is determined by the position of the retro-reflecting mirror. Electronic control of the optical lattice position, \emph{e.g.} for position stabilization or to modify the quantum properties of the system by applying a time-periodic force\cite{Arimondo2012,Lignier2007,Zenesini2008}, can be accomplished by moving the retro-reflected mirror. In the simplest approach using a mirror glued to a piezoelectric material, deformation of the piezo is not uniform and the mirror experiences a position dependent tilt. Typical path lengths in optical lattice experiments are fractions of a meter, and even a small amount of steering leads to a significant misalignment of the retro-reflected beam. This misalignment results in an unwanted modulation of the optical lattice depth, which motivated the present low-steering piezo mirror design.\\
\\
\indent The system is based on a triplet of individually controlled piezo actuators, which allows for independent adjustment of the expansion of each transducer in order to correct for imbalances between the three actuators.  The design substantially  reduces steering errors while maintaining a multi-micron displacement up to high frequencies. The system can vibrate $\varnothing=\SI{12.7}{\milli\meter}$ mirrors and is compatible with standard optical mounts. The easy to assemble design has been tested on an ensemble of atoms trapped in optical lattices.  This approach could be useful in a large set of applications, including high-finesse tunable optical cavities and live focus-stacking.


\begin{figure}[t]
    \centering
    \includegraphics[scale=0.2]{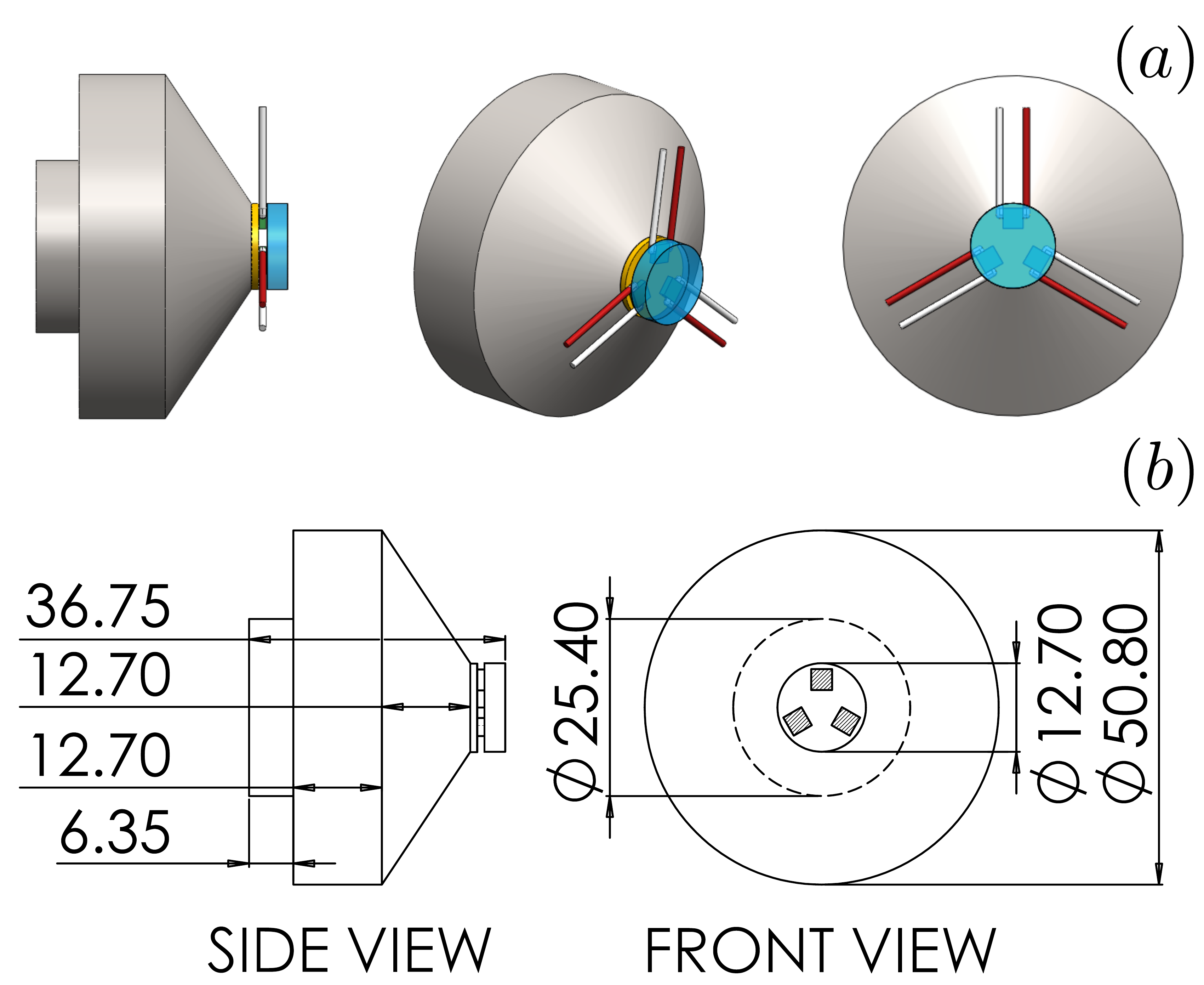}
    \caption{\textbf{Optomechanical design:} (a) The system includes a heavy steel support (metallic gray) compatible with standard $\varnothing=\SI{25.4}{\milli\meter}$ optical mounts, a ceramic electrical insulator (yellow), three piezo transducers (green with white and red cables) and a mirror (in blue). Technical plan of the system, all units in millimeter (b).}
    \label{solidworks}
\end{figure}
\section{Optomechanical design}
The present design is based on a triplet of piezo actuators (\textsf{Noliac NAC2012}\footnote{The identification of any commercial product or trade name does not imply endorsement or recommendation by the National Institute of Standards and Technology.}). These piezo plates ($3\times3$ \si{\milli\meter\squared}) combine a low capacitance (\SI{65}{\nano\farad}) with a relatively high maximum free stroke (nominally $\SI{3.3}{\micro\meter}$).  The piezos are epoxied onto the edges of a thin ceramic disk in an equilateral triangle (see Fig.\ref{solidworks}). The ceramic plate (alumina, $\SI{1}{\milli\meter}$ thickness, $\varnothing=\SI{12.7}{\milli\meter}$) facilitates electrical insulation between the piezos and the metal support. The plate is attached to a heavy steel support of mass $m=\SI{310}{\gram}$ which acts as a mechanical insulator, preventing resonances in the region of interest\cite{Briles2010}. While other materials (\emph{eg.} tungsten or marble) could be used, steel remains a good compromise between density, cost and machining time. The steel piece can be machined to fit within standard one-inch optomechanical mirror mounts. All surface bonds are performed with a slow-curing epoxy resin.\\  
\indent The design is compatible with $\varnothing=\SI{12.7}{\milli\meter}$ mirrors. Two types of mirrors have been tested: a $(R,e,\varnothing)=(\SI{-500}{\milli\meter},\SI{2}{\milli\meter},\SI{12.7}{\milli\meter})$ dielectric concave mirror (\textsf{Lattice Electro-Optics, Inc. RX-810-UF-MPC-0512-519}) and a $(e,\varnothing)=(\SI{3.1}{\milli\meter},\SI{12.7}{\milli\meter})$ flat silver mirror (\textsf{Newport ValuMax}). $R$ is the radius of curvature, $e$ the thickness and $\varnothing$ the diameter of the mirror. In the following, we refer to these two mirrors as $\mathcal{M}_{1}$ (concave mirror) and $\mathcal{M}_{2}$ (flat mirror). Section \ref{sec:performances} presents the performance of $\mathcal{M}_{2}$ while section \ref{sec:atoms} demonstrate the compatibility of $\mathcal{M}_{1}$ with an ensemble of cold atoms in optical lattices.


\section{Electronic system}
\begin{figure}[ht!]
    \centering
    \includegraphics[scale=1]{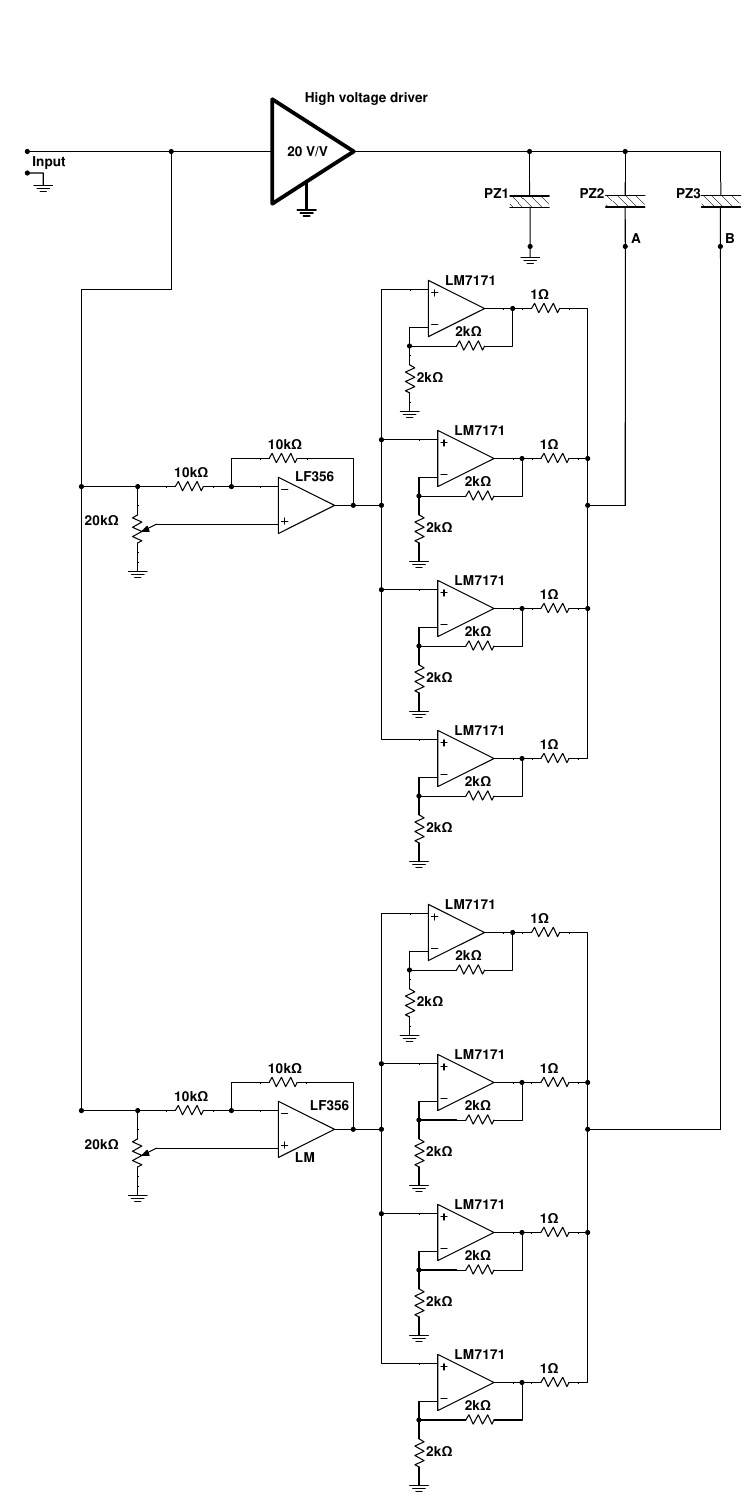}
    \caption{\textbf{Schematic of the piezo controller:} to improve readability the circuit diagram omits the power supply distribution and bypass network for the operational amplifiers described here briefly. All the operational amplifiers are powered from a dual-power supply between $\SI{\pm 15}{\volt}$. Each operational amplifier has an individual bypass capacitor (\SI{4.7}{\micro\farad}, \textsf{0805} multilayer ceramic) very close to the positive and negative supply pins. In addition a common $\SI{220}{\micro\farad}$ capacitor is placed between each voltage rail and ground. An effective bypassing of the power supply is critical to prevent instability especially when connecting several operational amplifiers in parallel.}
    \label{fig:schematicfig}
\end{figure}

The three piezo actuators are controlled by a tailor-made electronic device schematically shown on Fig.\ref{fig:schematicfig}. An external, commercial, ground-referenced high voltage driver (\textsf{PiezoDrive PX200}) provides most of the amplitude necessary to drive the piezo actuators. The driver has a gain of $20\si{\volt}/\si{\volt}$ over the control signal $V_{\text{Input}}$. The rest of the diagram shows the electronics used to vary the voltage across each piezo actuator. While PZ1 is referenced to ground, the other two piezos are referenced to nodes A and B. Through the $\SI{20}{\kilo\ohm}$ trim-pots the adjustment voltages $V_{\text{A}}$ and $V_{\text{B}}$ of the two nodes can be set to $V_{\text{A}}=k_{\text{A}}\cdot V_{\text{Input}}$ and $V_{\text{B}}=k_{\text{B}}\cdot V_{\text{Input}}$ where $k_{\text{A}}$ and $k_{\text{B}}$ can vary from $-2$ to $+2$. This allows for an overall gain of those two nodes to be adjusted over $\SI{18}{\volt}/\si{\volt}$ and $\SI{22}{\volt}/\si{\volt}$.\\
\indent The design of the second amplification stage of the driver is determined by the requirements of low dynamic output impedance and high output current capabilities that are needed when a highly capacitive load is driven at high voltage and high frequency. With driving frequencies of the order of $\SI{10}{\kilo\hertz}$ and peak amplitudes of the order of $\SI{100}{\volt}$
the peak current flowing in each piezo can be on the order of $(\SI{100}{\volt})\cdot2\pi\cdot(\SI{10}{\kilo\hertz})\cdot(\SI{65}{\nano\farad})=\SI{0.4}{\ampere}$.
\indent To obtain this current capability, four non-inverting gain stages based on a \textsf{Texas Instruments LM7171}\cite{LM7171_datasheet} are connected in parallel. The choice of this simple architecture is driven by the goals of limiting the component count, ensuring a flat frequency response over all the spectrum of operation, and eliminating the additional design complexity that a push-pull or totem-pole high-current output stage with discrete transistors usually introduces.
The \textsf{LM7171} has a current driving capability of $\SI{100}{\milli\ampere}$, therefore combining four amplification blocks in parallel allows to deliver to a load up to $\SI{400}{\milli\ampere}$ of peak current. In addition to the high current capability, the wide open-loop bandwidth of the operational amplifier ($\SI{120}{\mega\hertz}$) ensures a low dynamic output impedance over the entire frequency spectrum of operation of few tens of $\si{\kilo\hertz}$.\\
\indent When connecting in parallel multiple amplifiers, one issue is that small differences between their feedback networks cause them to compete with each other by forcing a slightly different voltage at the common output node. To prevent this undesirable behavior, a solution is to decouple the outputs by inserting small resistors in series. This potentially degrades the performance by increasing the overall output impedance but in this particular circuit the value of just $\SI{1}{\ohm}$ shown in Fig.\ref{fig:schematicfig} is sufficient to decouple the four amplifiers. The additional increase in the output impedance is only $\SI{0.25}{\ohm}$. All \textsf{LM7171} are configured as non-inverting buffers with a gain of 2.\\
\indent When holding a DC output voltage, or for AC drives during the typical cold atoms experimental timescales ($\tau \leq \SI{100}{\milli\second}$), the \textsf{LM7171} do not require heat sinking. Full schematics, printed circuit board artwork and bill of material for the fabrication of the circuit in Fig.\ref{fig:schematicfig} are available on the JQI git repository \cite{git}.


\section{Performance}
\label{sec:performances}
In this section, we characterize the mechanical response, the maximum stroke, the static steering-error and the amplitude of the dynamic tilt. These measurements have been performed with $\mathcal{M}_{2}$. 

	\subsection{Frequency response}
\begin{figure*}[ht!]
    \centering
    \includegraphics[width=1\textwidth]{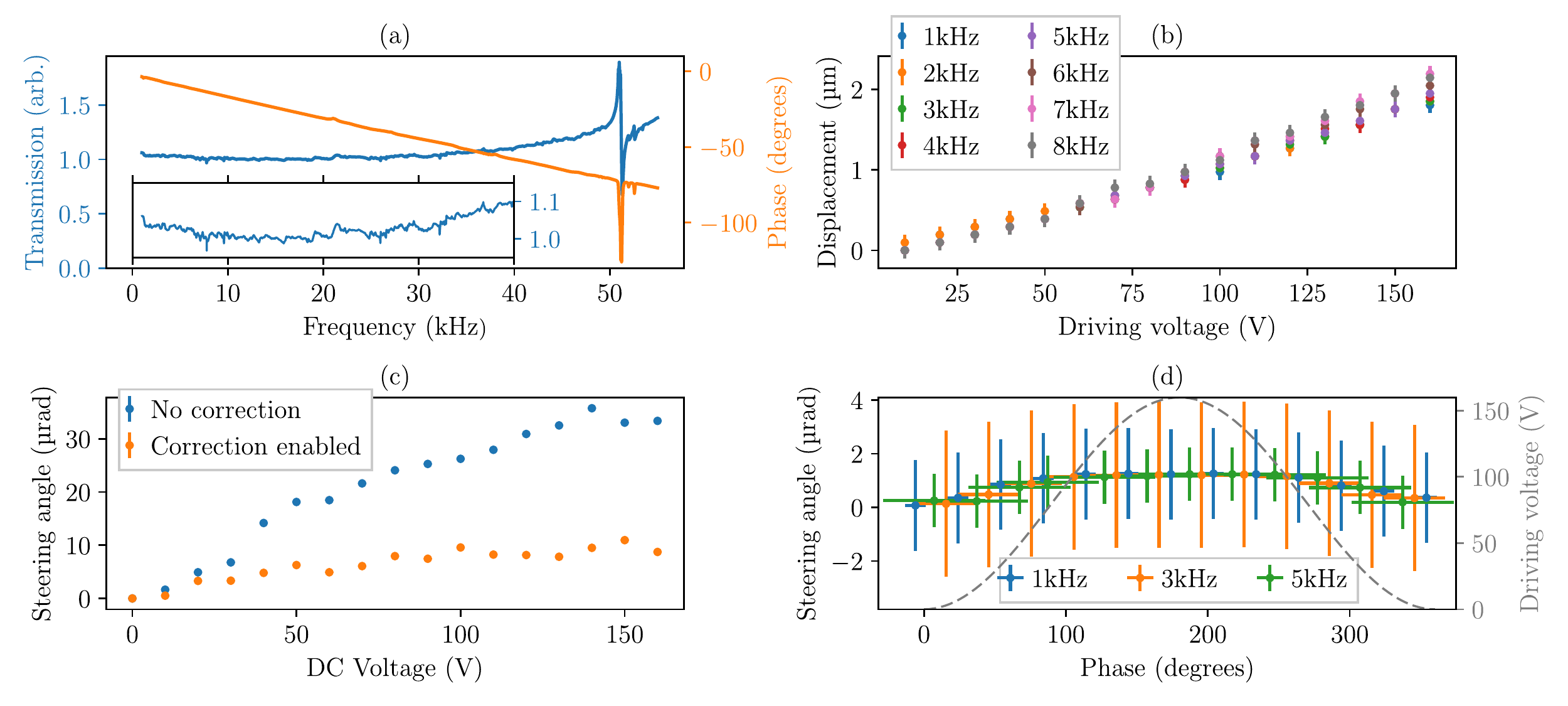}
    \caption{\textbf{Characterization of the system:} (a) The frequency response (gain in blue, phase in orange) is flat up to $\SI{40}{\kilo\hertz}$. In inset, we plot a zoom of the transmission in the region of interest (\SI{1}{\kilo\hertz} to \SI{40}{\kilo\hertz}, $\times 10$ magnification on the $y$-axis). (b) We achieve a $\SI{2}{\micro\meter}$ displacement up to $\SI{8}{\kilo\hertz}$. (c) The feed-forward systems enables to gain a factor 3.5 on the static steering error. (d) We observe a maximum dynamic steering error lower than $\SI{0.6}{\micro\radian\micro\meter}^{-1}$. The uncertainty of the phase is due to the exposure time of the beam profiler. The dashed gray line indicates the instantaneous voltage. On both (c) and (d), the vertical error bars correspond to the standard deviation of the fit to a gaussian.}
    \label{freqfig}
\end{figure*}

We measure the frequency response of the piezo-tripod by interferometry. The setup is based on a Michelson interferometer set to measure the optical path length of a vibrating arm. The interferometer has an arm length about $l\approx\SI{100}{\milli\meter}$ and uses a collimated beam from a $\lambda=\SI{780}{\nano\meter}$ laser diode.\\ 
\indent In order to compensate slow drifts, we phase-lock the test interferometer. The reference arm is equipped with a separate piezo-actuated mirror controlled by a feedback loop. This corrects low frequency drifts due to thermal and mechanical imperfections. The locking frequency of the PID loop is $f_{\text{lock}}\approx\SI{100}{\hertz}$, which is an order of magnitude lower than the smallest frequency tested on the piezo-tripod.\\
\indent A periodic path length difference $\Delta x$ results in a modulation of the intensity $\Delta I$. Around half-maximum, the $\sin^{2}$ dependence of the intensity in $\Delta x$ can be linearized. In this region and assuming small path length differences ($\Delta x \ll \lambda/2$), the variation of intensity becomes proportional to the mirror displacement, $\Delta I \propto \Delta x$.\\
\indent We use a \textsf{Bode 100} network analyzer to generate a frequency sweep from $\SI{1}{\kilo\hertz}$ to $\SI{55}{\kilo\hertz}$ which is then amplified and sent to the piezo tripod. The modulation of intensity is measured by a photodiode and separated between the PID feedback loop and the network analyzer input. We observe a flat response up to $\SI{40}{\kilo\hertz}$ (see Fig.\ref{freqfig}(a)). The phase is decreasing linearly with a remarkable absence of resonances up to $\SI{40}{\kilo\hertz}$. We attribute the linear decrease to a large mechanical resonance at higher frequencies.

	\subsection{Maximum displacement}
The maximum displacement characterization involves the same interferometer, with the phase-locking piezo of the reference arm disabled. The piezo-tripod is driven with a periodic symmetric triangular signal, leading to a time-dependent path length difference $2\delta(t)$ with the same characteristics. Over one period, we observe $n$ minima of intensity from which we can recover the mirror displacement $\delta$. With the given drive circuit operating at maximum amplitude, we achieve displacements up to $\SI{2}{\micro\meter}$ for frequencies between $\SI{1}{\kilo\hertz}$ and $\SI{8}{\kilo\hertz}$ (see Fig.\ref{freqfig}(b)). 

	\subsection{Static steering error}	
We characterize the amplitude of the static steering error by measuring the lateral displacement of a retro-reflected beam at large distance. We place a beam profiler \textsf{Thorlabs BC106N-VIS} at $D=\SI{12.38\pm.03}{\meter}$ from the vibrating mirror. We take a reference position $\mathbf{r}_0=\mathbf{r}(V=0)$. We then apply a DC voltage $V$, measure the position of the reflected beam $\mathbf{r}(V)$ and determine the tilt angle $\theta(V)=\arctan[(r(V)-r_{0})/D]$.\\
\indent For each position, we average a set of 10 images and extract the position of the center with a two-dimensional gaussian fit. The exposure time is $\SI{1}{\second}$, which helps to filter vibrational noise. The images are taken after thermalization of the optomechanics.\\
\indent We first measure the maximum static tilt without using the adjustment trim-pots of the electronic board, $k_{A}=k_{B}=0$. We observe a $\SI{17.9(3)}{\micro\radian\micro\meter}^{-1}$ tilt. With the trim-pots adjusted, this reduces to $\SI{5.5\pm0.3}{\micro\radian\micro\meter}^{-1}$ (see Fig.\ref{freqfig}(c)). We attribute the residual tilt to low-frequency thermal drifts. The instrinsic hysteresis of each piezo may also be detrimental to pointing stability. The same experiment performed with a single, center-mounted $10\times10\, \si{\milli\meter}^{2}$ piezo plate (\textsf{Noliac NAC2015}) leads to a static tilt larger than $\SI{250}{\micro\radian\micro\meter}^{-1}$. We note that even the uncompensated design improves the steering over a single-piezo design. With drive compensation, the steering of the tripod design is $50$ times smaller. 

	\subsection{Time-dependent residual tilt}
The time-dependency of the tilt with a periodic shaking is measured via a similar technique. We drive the piezo-tripod with a sinusoidal signal at frequency $f$. The beam profiler is now placed at $D=\SI{7.76\pm 0.03}{\meter}$ from the mirror and is triggered with a pulse signal at the same frequency $f$ than the drive. The drive phase $\phi$ sets the instantaneous voltage at which the picture is taken.\\
\indent In order to limit effects due to heating, we gate the two signals with a $\SI{10}{\hertz}$ square signal, so that we only shake ten periods every $\SI{100}{\milli\second}$. For each trigger phase, we extract the position of the retro-reflected beam from an average over 20 fitted images. Each image has an exposure time of $\SI{20}{\micro\second}$, which is short enough to resolve frequency-dependent tilts up to $\SI{5}{\kilo\hertz}$.  The uncertainty of the measured position is given by the fit residuals.\\
\indent We measure a maximum dynamic tilt of $\SI{1.25}{\micro\radian}$ for three different frequencies ($1,3$ and 5\,\si{\kilo\hertz}) shaken at maximum amplitude ($\SI{160}{\volt}$), see Fig.\ref{freqfig}(d). This largely reduced steering error, by a factor of 10 compared to the static case, suggests that most of the DC steering is due to low frequency thermal effects. The dynamic tilt of the compensated tripod design of $\SI{0.6}{\micro\radian\micro\meter}^{-1}$ is 17 times smaller than the single piezo design. 

\begin{figure*}[ht!]
    \centering
    \includegraphics[width=1\textwidth]{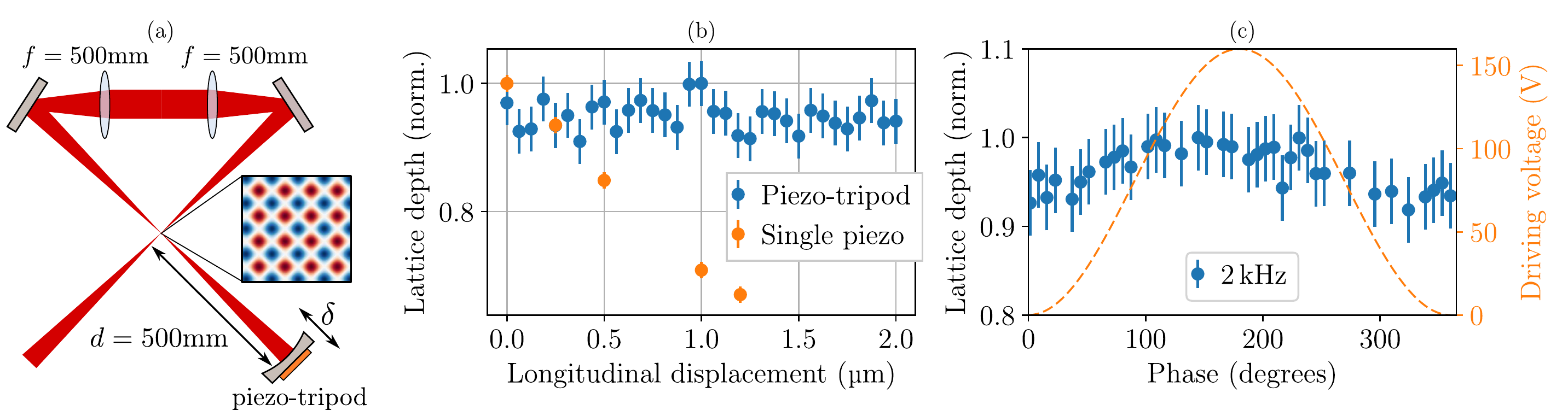}
    \caption{\textbf{Shaking an optical lattice:} (a) The bowtie-shaped optical path leads to a two-dimensional optical lattice at the crossing of the beams. (b) We compare the performances of a single piezo actuator (\textsf{Noliac NAC2015}) with the piezo-tripod for DC voltages. While the first becomes rapidly unusable, the latter performs well up to its maximum displacement. The errorbars of the piezo-tripod measurement are given by $F$, the ones of the single-piezo data correspond to $F/\sqrt{10}$. (c) We measure a lattice depth modulation lower than $8\,\%$ for $f=\SI{2}{\kilo\hertz}$, with errorbars equal to $F$. Identical measurements done at $1,3$ and $\SI{4}{\kilo\hertz}$ lead to similar results.}
    \label{fig:atoms}
\end{figure*}


\section{Application : shaking cold atoms in optical lattices} \label{sec:atoms}
Atoms in optical lattices are extremely sensitive to the depth of the trapping potential\cite{Bloch2005a}. In the case of a bowtie-shaped lattice\cite{Sebby-Strabley2006}, misalignments in the retro-propagation as small as a few tens of microradians lead to a visible degradation of the trapping depth.\\
\indent Our experimental setup consists of a $^{87}\text{Rb}$ Bose-Einstein Condensate of $N\approx 4\times10^{4}$ atoms and typical dimension $a=\SI{10}{\micro\meter}$ loaded into a bowtie-shaped lattice made of $\lambda=\SI{813}{\nano\meter}$ light. $\mathcal{M}_{1}$ is closing the bowtie and is placed at $d=\SI{500}{\milli\meter}$ from the atomic cloud (see Fig.\ref{fig:atoms}(a)). This leads to a $1/e^{2}$ beam radius of $\sigma\approx\SI{170}{\micro\meter}$ on the atoms.\\
\indent The optical lattice depth can be measured with the atoms by Kapitza-Dirac diffraction\cite{Gould1986}, in which the lattice light is switched on and off and the lattice depth is determined from the resulting atom diffraction.\\
\indent To test the alignment for constant voltages, we step the position of the piezo and observe the evolution of the lattice depth. Over a $\delta=\SI{2}{\micro\meter}$ displacement, the deterioration of the lattice depth is lower than the noise floor ($F=\pm 4\, \%$). The same experiment made with a single $\SI{100}{\milli\meter}^2$ piezo-plate (\textsf{Noliac NAC2015}) leads to a reduction of the lattice depth larger than $30\%$ over less than a $\SI{1.2}{\micro\meter}$ (see Fig.\ref{fig:atoms}(b)). For this latter experiment, each datapoint corresponds to an average of $10$ measurements.\\ 
\indent To measure the instantaneous steering error, we set the mirror in a sinusoidal motion at maximum driving amplitude. We then send the pulse of lattice light ($\SI{2}{\micro\second}$ duration) at different phases of the drive and extract the lattice depth from the diffraction pattern. Up to $\SI{4}{\kilo\hertz}$, the measured variation in the lattice depth is lower than $8\, \%$ (see Fig.\ref{fig:atoms}(c)).\\


\section{Conclusion} \label{sec:conclusion}
We present the design of a retro-reflecting vibrating mirror combining high frequencies, high amplitude and low steering error. Along with its relatively low cost, the system combines efficiency and compactness. It allows to vibrate relatively large mirrors ($\varnothing=\SI{12.7}{\micro\meter}$) with a flat frequency response, in principle up to tens of \si{\kilo\hertz} and fits into standard optomechanical mounts.\\
\indent While this design was originally optimized for optical lattices, its versatility could be applicable for a wide range of applications, such as high-finesse tunable optical cavities, interferometric microscopy and laser stabilization. \indent We believe that further modifications could possibly reduce the steering error, \emph{e.g.} increasing the number of piezo actuators or driving each piezo with an independent waveform.\\ 

\begin{acknowledgments}
We thank T. Esslinger and his team for useful suggestions and for providing some of the early experimental tools and insights. This work was partially supported by ARL-CDQI and NSF PFC at JQI. E.M. acknowledges the support of the Fulbright program and NSF PFC. T.B. acknowledges the support of the European Marie Sk\l{}odowska-Curie Actions (H2020-MSCA-IF-2015 Grant 701034).
\end{acknowledgments}

\bibliography{Tripodv7}

\end{document}